\documentclass{article}
\usepackage{graphicx}
\usepackage{amsmath,amssymb}
\usepackage{color}
\usepackage{natbib}

\begin{document}

\begin{center}
\large
\textbf{Measurement of the pure dissolution rate constant of a mineral in water} \\
\vspace{0.5cm}
\normalsize
\textsc{Jean Colombani\footnote{Jean.Colombani@univ-lyon1.fr}}\\
\small
Laboratoire de Physique de la Mati\`ere Condens\'ee et Nanostructures,\\
Universit\'e de Lyon, F-69003, Lyon, France;\\
Universit\'e Claude Bernard Lyon 1, F-69622 Villeurbanne, France;\\
CNRS, UMR 5586, F-69622 Villeurbanne, France
\normalsize
\vspace{1cm}

Revised version - 5 september 2008
\vspace{1cm}

\begin{minipage}{15cm}\textbf{Abstract---}
We present here a methodology, using holographic interferometry,
enabling to measure the pure surface reaction
rate constant of the dissolution of a mineral in water, unambiguously free from
the influence of mass transport.
We use that technique to access to this value for gypsum and we demonstrate that
it was never measured before but could be deduced a posteriori from the
literature results if hydrodynamics is taken into account with accuracy.
It is found to be much smaller than expected.
This method enables to provide reliable rate constants for the test of dissolution models
and the interpretation of in situ measurements, and gives clues to explain the inconsistency between dissolution rates of calcite and aragonite, for instance,
in the literature.
\end{minipage}
\end{center}

\section{INTRODUCTION}

Among heterogeneous reactions, dissolution of minerals in water is
encountered in a wide spectrum of fields, from geochemistry to materials
science, soil science, environmental science or oceanography.
It plays a leading role, for instance, in the weathering of rocks
as much as in the durability of mineral materials, in soils amendment,
in pollutant spreading, or in sediment-water interaction.
In all these situations, quantitative kinetic models are required to describe the
involved phenomena.

More recently, this phenomenon has been recognized as being a keypoint
in the modelisation of the atmospheric $P_{\mathrm{CO}_2}$ variation.
Indeed, among the mechanisms controlling the pH of the ocean,
and influencing \textcolor{green}{in turn} the atmospheric $P_{\mathrm{CO}_2}$, dissolution drives
two major steps: the weathering of terrestrial carbonates
and the deep-sea sediment-water interaction \citep{Anderson}.
Therefore kinetic models of $P_{\mathrm{CO}_2}$ change require reliable
in situ and laboratory dissolution rates \citep{Hales}.

But the establishment of reliable rate laws faces several difficulties.
First, multiple lengthscales, and accordingly timescales, are involved
during the dissolution process of a mineral, from the individual
atomic detachment to the change of shape of the solid.
Secondly, the number of parameters influencing the global dissolution kinetics
is particularly large: surface topology, amount and nature of surface defects,
adsorbed species, reactive surface area, pressure, temperature,
composition of the solution, hydrodynamic behavior of the liquid, pH, \dots
So a comprehensive dissolution model is still lacking and experimental results
of dissolution rates show inconsistencies \citep{Morse}.

To obtain rate laws of the dissolution of minerals in water, two methodologies
are employed.
In bulk experiments, the concentration of chemical species in stirred water,
where a sample dissolves, is measured
and the dissolution rate deduced from the time evolution of this concentration.
In local experiments by Atomic Force Microscopy \textcolor{green}{(AFM)} or Vertical Scanning
Interferometry \textcolor{green}{(VSI)}, the evolution of the surface topology at the molecular scale
is followed during dissolution and the kinetics possibly deduced from atomic step velocity
or surface-normal retreat measurement \citep{Vinson}.
Agreement between bulk and local techniques remains scarce
(cf. \cite{Arvidson} for the case of calcite, \cite{Luttge} for dolomite \textcolor{green}{,
\cite{Luttge06} for albite and \cite{Jordan} for magnesite).}

In both experiments the solvent is flowing past the mineral surface,
in a laminar or turbulent regime.
A mass transport
boundary layer develops in the vicinity of the solid, in which the
liquid velocity grows from zero at the interface (the no-slip condition), to its bulk value.
In this context, dissolution proceeds through three steps (cf. Fig. \ref{scheme}).
First the ions are unbound from the solid and solvated.
Following a transition state theory the kinetics of the matter removal from the surface
obeys a power law $R_{\mathrm{diss}}=k_{\mathrm{s}}(1-c_{\mathrm{s}}/c_{\mathrm{sat}})^n$,
where $R_{\mathrm{diss}}$
is the dissolution rate, $k_{\mathrm{s}}$ the surface reaction rate constant,
$c_{\mathrm{s}}$ the concentration of the dissolved species at the surface,
$c_{\mathrm{sat}}$ their solubility and $n$ a constant \citep{Lasaga98}.
Then the ions migrate through the diffusional boundary layer (DBL).
The concentration is generally considered as linear in this layer and Fick's law writes
$R_{\mathrm{diff}}=k_{\mathrm{t}}(c_{\mathrm{s}}-c)$, where $R_{\mathrm{diff}}$ is the diffusion rate,
$k_t=D/\delta$ a transport rate constant, $D$ the diffusion coefficient of the dissolved
components, $\delta$ the DBL thickness and $c$ the concentration in the bulk liquid \citep{Lasaga98}.
Finally the ions are advected by the flow toward the concentration measurement device.
The kinetics of the phenomenon is driven by the slowest step, so minerals
are classified according to their transport-controlled (e.g. rock salt), when
$R_{\mathrm{diff}}\ll R_{\mathrm{diss}}$,
or reaction-controlled (e.g. quartz), when
$R_{\mathrm{diff}}\gg R_{\mathrm{diss}}$, character.
Carbonate and sulfate minerals are considered to belong to a mixed kinetics class
where both the transport and reaction flow rates are comparable \citep{Jeschke,Rickard}.

When mass transport is identified as the limiting step,
the pure dissolution coefficient $k_{\mathrm{s}}$ is inaccessible and only
$k_t=D/\delta$ is obtained in bulk experiments.
Though this transport coefficient has been measured in numerous dissolution studies
and is sometimes relevant for the understanding of
dissolution scenarii in the field or in industrial situations, it does not concern,
from a conceptual point of view, the nature of the mineral and the physicochemistry of its
surface:
the boundary layer depth $\delta$ is a pure hydrodynamical quantity and the diffusion coefficient
value of the common ions is always $D\sim$10$^{-9}$ m$^2$ s$^{-1}$.
For mixed kinetics studies, also largely common,
with the help of the above laws, an empirical surface rate constant
$k$ can be obtained  "function of both surface reaction and mass transport"
\citep{Jeschke}, which is different from $k_s$,
whereas only this coefficient contains the chemistry of water-mineral
interactions and varies by orders of magnitude from minerals to others.

In this paper, (i) we recall that
the kinetics of matter removal during dissolution is characterized by
a pure surface reaction rate constant, independent
of the concentration \textcolor{green}{field in the liquid}
 and \textcolor{green}{of the transport kinetics} from the surface to the bulk liquid, and
(ii) we claim that this pure dissolution rate constant can be measured unambiguously,
even for transport-controlled and mixed kinetics,
provided that the concentration field above the surface is known.
To ascertain this claim, we have followed the subsequent procedure.
First a working mineral and a relevant experimental technique have been selected
(Section \ref{material}).
Then reliable measurements of the pure surface reaction rate constant of this mineral
have been performed in conditions where mass transport is proven to have no influence
(Section \ref{results}).
Subsequently most of the experimental dissolution rates available in the literature
have been critically collected (Section \ref{literature}).
Finally the validity of these measurements and of our results is discussed (Section \ref{discussion})

\section{MATERIAL AND METHOD}
\label{material}

To be able to measure the pure dissolution coefficient without ambiguity, i.e.,
without needing hydrodynamical assumptions, we need (i) to observe dissolution
in absence of any convective flow and (ii) to access to the concentration field in
the liquid soaking the solid.
To achieve these two goals, we have carried out holographic interferometry
measurements of the dissolution of a gypsum single crystal in water at rest.

\textcolor{green}{Holography proceeds in two steps.
First the hologram of an object is recorded by enlightening a photographic plate
with both the beam coming from the object and a known reference beam.
Whereas a classical photograph, where the plate is only enlightened by the object beam,
exclusively contains informations about the light amplitude,
the interference pattern of the object and reference beams contains informations
about both the amplitude and the phase.
The phase itself is linked to the optical path, so to the third dimension and to
the refractive index of the object.
Secondly the hologram is enlightened only by the reference beam.
This one is diffracted by the recorded interference pattern, which gives birth
to a three-dimensional image of the object.}

Here we use the potentiality of holography to record phase objects, i.e., transparent
objects exhibiting only variation of their index of refraction (a transparent liquid, in our case).
We record first the three-dimensional state of the studied system in a hologram at time $t_0$.
Subsequently we enlighten both the hologram with the reference beam ---which creates a 3D image
of the object at $t_0$--- and the object.
Thereby we have a superposition of the object at time $t$ and of this object at time $t_0$.
If a variation has occurred between these two times, of concentration for instance, it is
visualized through interference fringes.
Holographic interferometry differentiates from classical interferometry in the fact that no external reference
is needed, the object interfering virtually with a memory of itself.

\begin{figure}
\begin{center}
\includegraphics[width=0.6\linewidth]{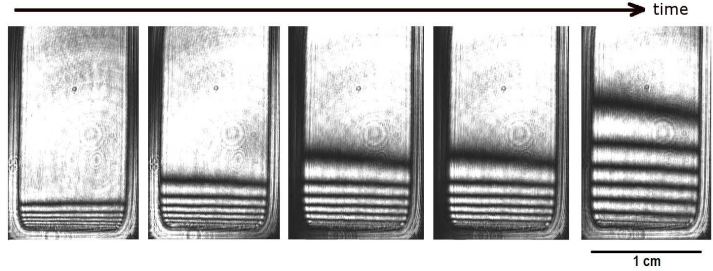}
\caption{Interferograms of the dissolution of a Mazan gypsum single crystal
20, 65, 105, 200 and 377 min. after the start of the experiment.
The crystal is a thin grey line at the bottom of the optical cell.}
\label{holog}
\end{center}
\end{figure}

In the course of an experiment, the hologram of a thermostated optical cell filled
with pure water is registered.
Then a thin mineral sample is introduced in the cell at a time considered as the time
origin $t_0$.
Subsequently both the object and the hologram are enlightened and their
interference pattern is recorded periodically as shown in Fig. \ref{holog}.
As the mineral dissolves the concentration of the dissolved species in water evolves,
so does the refractive index, and the resultant optical path length difference between the
object at time $t$ and time $t_0$ (recorded in the hologram) induce interference fringes.
From the topography of these fringes, the two-dimensional concentration field in the cell
can be deduced.
Parallely second Fick's law has been resolved with a first order chemical reaction
at the solid-liquid interface
in a semi-infinite one-dimensional approximation and brings the concentration evolution
$c(z,t)$ with vertical dimension $z$ and time $t$.
The best fit of the experimental curves deduced from holographic results with this analytical
expression brings the pure surface reaction rate constant $k_s$.
Details on the experimental setup, procedure and data analysis have been given elsewhere \citep{Colombani07}.

To validate our assertion \textcolor{green}{on the hydrodynamical bias present in classical
dissolution measurements}, we need a mineral with the following requirements.
First its dissolution rate must give a discernable amount of dissolved matter
in laboratory times.
Secondly, its kinetics must be either transport-controlled or mixed.
Thirdly the chemical reaction must be as simple as possible, to focus
on the hydrodynamical and topological aspects of dissolution.
We have therefore discarded calcite, although most effort on mineral dissolution in water
has concerned this mineral, because the various steps of the reaction
make its kinetic law complex and strongly dependant on the pH and
$P_{\mathrm{CO}_2}$ values
during the experiments.
Gypsum (CaSO$_4$, 2H$_2$O) has been chosen because it fulfills these three criterions
and because numerous literature results are available.
Beside these methodological considerations, we must notice that gypsum dissolution
liberates Ca$^{2+}$, a cation liable
to fix CO$_2$, and is consequently of importance for the geological sequestration of this
greenhouse gas.

\section{RESULTS}
\label{results}

\textcolor{green}{We had measured the dissolution coefficient of the cleavage plane of
gypsum from the Mazan quarry (France), in the frame of a study presenting the possibility
of performing mineral dissolution measurement by holographic interferometry \citep{Colombani07}.
This work was focussed on the experimental protocole and the cleaved Mazan gypsum sample was studied
as an example.}
Our primary goal \textcolor{green}{here} is the demonstration of the experimental bias introduced by the liquid flow
in the classical dissolution experiments (by solution chemistry or local probe).
But one difficulty to confront the literature measurements and ours
lies in the fact that the mineral geometrical and chemical properties differ from one setup
to another.
In global studies, samples are \textcolor{green}{often powders, sometimes compacted, or polished crystals,}
whereas we use cleaved single crystals.
So when only one identified interface dissolves in our case, dissolution in literature stems
from multiple crystalline planes separated by steps and kinks.
Furthermore the gypsum origin (so the impurities) differ between experiments.
\textcolor{green}{The only sample studied in our preceding work is therefore not sufficient
to estimate the dispersion induced by this variation of gypsum samples.}
To capture the influence of these
differences on $k_s$ we have compared some samples to our reference one, i.e., the (010)
cleavage plane
of gypsum from the Mazan quarry at 20.00$^{\circ}$C
($k_{\mathrm{s}} = (4\pm 1)\times$10$^{-5}$ mol m$^{-2}$ s$^{-1}$) \citep{Colombani07}.
We have carried out holographic interferometry experiments of the dissolution of
gypsums of two other quarries: Tarascon-sur-Ari\`ege in France
($k_{\mathrm{s}} = 4\times$10$^{-5}$ mol m$^{-2}$ s$^{-1}$) and
Mostaganem in Algeria ($k_{\mathrm{s}} = 7\times$10$^{-5}$ mol m$^{-2}$ s$^{-1}$)
and a synthetic commercial one from MaTeck GmbH ($k_{\mathrm{s}} = 3\times$10$^{-5}$ mol m$^{-2}$ s$^{-1}$),
to check the influence of the origin.
Then we have carried out experiments with the (120) plane of Mazan gypsum
polished with silicon carbide paper of grit size down to 15 $\mu$m
($k_{\mathrm{s}} = 7\times$10$^{-5}$
mol m$^{-2}$ s$^{-1}$), to evaluate the behavior of non-cleaved planes.

\textcolor{green}{All of these measurements deserve their own study and they were performed here only to
estimate the variability of $k_{\mathrm{s}}$ in various experimental conditions.
But, as a first approach, we can notice that the dispersion among the $k_{\mathrm{s}}$ of the (010) samples
of various origins ($k_{\mathrm{s}} = 3$ to $7\times$10$^{-5}$ mol m$^{-2}$ s$^{-1}$)
is greater than
the standard error deduced from seven measurements on the (010) Mazan sample in our preceding work
($\Delta k_{\mathrm{s}}=1\times$10$^{-5}$ mol m$^{-2}$ s$^{-1}$).
The choice of the quarry has therefore a non negligible influence on the dissolution rate constant.
The (010) and (120) planes have different chemical and topological characters.
The first one shows originally only water molecules and tends to be atomically flat \citep{Fan}
whereas the second one is rough and exhibits water molecules, cations and anions.
But despite these differences, there is less than a factor of two between their dissolution coefficients.}

\textcolor{green}{We are just searching for a $k_{\mathrm{s}}$ range representative of the
$k_{\mathrm{s}}$ encountered in all the samples of the literature (cf. Section \ref{discussion})
so no average value was computed.}
Instead, if $k_{\mathrm{s}}^{\mathrm{max}}$ and $k_{\mathrm{s}}^{\mathrm{min}}$
are the maximum and minimum values, the nominal value of $k_s$ has been chosen as
$(k_{\mathrm{s}}^{\mathrm{max}}+k_{\mathrm{s}}^{\mathrm{min}})/2=5\times10^{-5}$ mol m$^{-2}$ s$^{-1}$
with an uncertainty of
$(k_{\mathrm{s}}^{\mathrm{max}}-k_{\mathrm{s}}^{\mathrm{min}})/2=2\times10^{-5}$ mol m$^{-2}$ s$^{-1}$.

\section{LITERATURE ANALYSIS}
\label{literature}

\begin{figure}
\begin{center}
\includegraphics[width=0.7\linewidth]{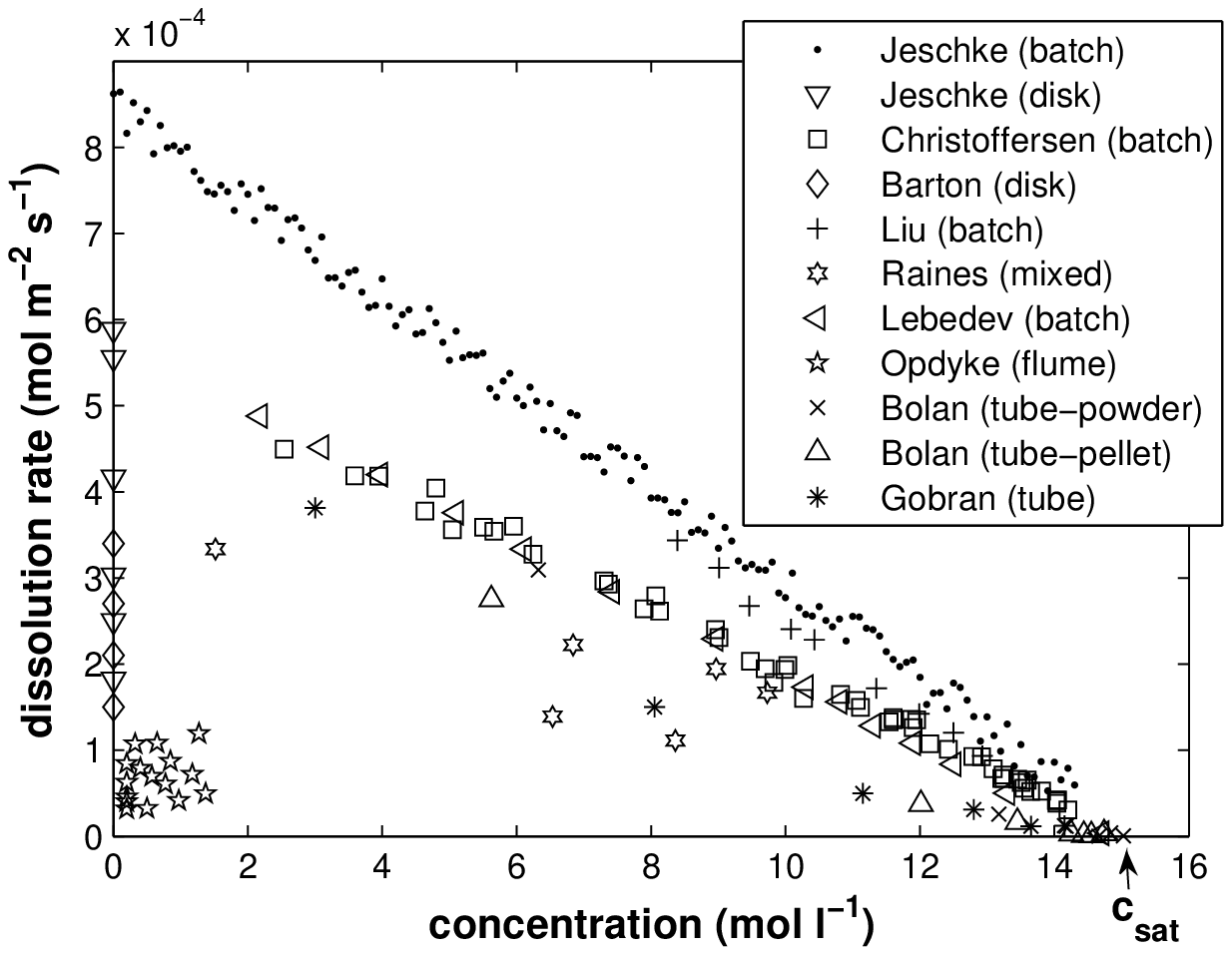}
\caption{Dissolution rates $R$ of gypsum in water as a function of calcium
concentration $c$ for various setups.
The $R(c)$ curves are drawn from stirred batch experiments by
\cite{Jeschke}, \cite{Christoffersen}, \cite{Liu} and \cite{Lebedev}, from rotating disk
experiments by \cite{Jeschke} and \cite{Barton},
from mixed flow/rotating disk reactor experiment by \cite{Raines},
from shaken tube experiments by \cite{Bolan} (with powder or
compressed powder pellets) and \cite{Gobran} and flume experiments
by \cite{Opdyke}.
$c_{\mathrm{sat}}$ is the solubility of the dissolved calcium ions in water.}
\label{liteRc}
\end{center}
\end{figure}

Dissolution rates of gypsum in water have been measured for various purposes
(karst formation, soil amendment, marine biology, \dots) and we have tried
to gather and analyse these measurements to obtain a coherent view of the literature data.
Fig. \ref{liteRc} collects the dissolution rates as a function of calcium concentration
of most of the literature results.
These include batch (dissolution in a reactor with stirrer),
rotating disk, shaken tube, mixed reactor
(rotating disk + flowing liquid) and flume experiments.
Unfortunately flowing cell results could not be exploited because dissolution occuring
all along the cell, a fixed concentration could not be ascribed to the value
of the dissolution rate at the output of the cell \citep{Kemper,Keren}.
Gypsum dissolution has been also widely used in marine biology to estimate
biological objects interaction with water.
But as was extensively demonstrated by \cite{Porter}
hydrodynamic conditions are often badly or wrongly defined in these studies and
dissolution rates cannot be ascribed to unamibiguous concentration and DBL thickness values.

\textcolor{green}{In all these studies, the solvent is pure water and
the working temperature is either 20 or 25$^{\circ}$C.
The pH and ionic strength values are almost never given.
But despite this lack, the gypsum solubility obtained in all these works is always
$c_{\mathrm{sat}}\approx 15$ mmol l$^{-1}$, identical to the reference value deduced
from literature analysis by \cite{Christoffersen} or \cite{Raju}.
This proves the proximity between the chemical
conditions of these experiments, and gives confidence in the possibility
of comparing their dissolution rates.}

\begin{figure}
\begin{center}
\includegraphics[width=0.6\linewidth]{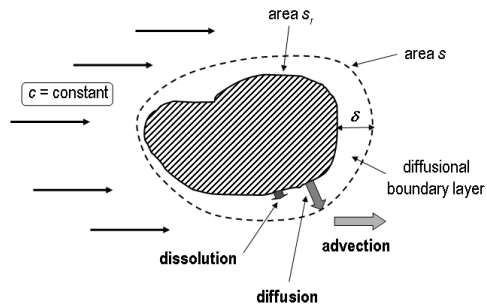}
\caption{Scheme of the successive mechanisms occuring during the dissolution of a crystal grain in a bulk solution chemistry experiment.}
\label{scheme}
\end{center}
\end{figure}

At first glance, all these measurements do not provide a consistent description
of the dissolution of gysum in water (cf. Fig. \ref{liteRc}).
Up to now, attention has been paid before all to the exponent of the power law of $R(c)$
---considered as representative of the kinetics---, the prefactor being ascribed
to the specificities of each experiment.
But a correct consideration of the hydrodynamical condition of each device
should bring a consistent and predictive knowledge of the dissolution process,
whatever the experimental setup.
So, first of all, we do not postulate that experiments are either transport-controlled of
reaction-controlled and we make the assumption that all of them are
influenced by both chemical reaction and
diffusive transport in a variable proportion.
The systems are always carefully maintained in a quasi-steady state,
where the concentration is uniform in the whole bulk liquid and evolves slowly.
So keeping in mind the dissolution steps shown in Fig. \ref{scheme},
we consider that matter is conserved during migration in the diffusional
boundary layer.
We apply mass conservation between the bottom and top of the DBL,
equalizing the dissolution flow rate $s_{\mathrm{r}}R_{\mathrm{diss}}$ and diffusion
flow rate $sR_{\mathrm{diff}}$:

\begin{equation}
sD\frac{c_{\mathrm{S}}-c}{\delta}=s_{\mathrm{r}}k_{\mathrm{s}}(1-\frac{c_{\mathrm{s}}}{c_{\mathrm{sat}}})
\end{equation}

with $s_{\mathrm{r}}$ the reacting surface area and $s$ the outer surface area of the DBL
(similar to the so-called geometric surface area of the solid, cf. Fig. \ref{scheme}).
We have followed \cite{Jeschke} and considered that $n$=1 in $R_{\mathrm{diss}}$.
From this equation, we compute the only unknown parameter, the surface concentration
$c_{\mathrm{s}}$,
and we can write the experimental dissolution rate $R=R_{\mathrm{diff}}$ measured
far from the solid after being advected:

\begin{equation}
R=\frac{1}{\frac{s}{s_{\mathrm{r}}k_{\mathrm{s}}}+\frac{\delta}{Dc_{\mathrm{sat}}}}
-\frac{1}{\frac{sc_{\mathrm{sat}}}{s_{\mathrm{r}}k_{\mathrm{s}}}+\frac{\delta}{D}}c
\label{R_c}
\end{equation}

The $\delta=0$ situation reflects an hypothetical absence of DBL where the kinetics
is directly driven by the reaction, the transport time between the surface and
the measurement device becoming negligible.
We see that the pure surface reaction rate constant $k_{\mathrm{s}}$ can be accessed via
$\lim_{\delta\rightarrow 0}1/(\partial R/\partial c)=sc_{\mathrm{sat}}/(s_{\mathrm{r}}k_{\mathrm{s}})$.
For this purpose, we need an evaluation of $c_{\mathrm{sat}}$, $\delta$ and $s/s_{\mathrm{r}}$.
The solubility is known for long: $c_{\mathrm{sat}}=15$ mmol~l$^{-1}$ \citep{Raju}.
The diffusional boundary layer thickness
wrapping the dissolving solid for each device is more delicate to obtain
This quantity is clearly identified in rotating disk experiments:
$\delta=1.61D^{1/3}\nu^{1/6}\omega^{-1/2}$ with
$\nu$ the kinematic viscosity of the solution and $\omega$ the angular
velocity of the disk \citep{Barton}.
For batch experiments, we have considered that at first order the liquid
develops a layer as if flowing past a semi-infinite plate:
$\delta=(2\pi/0.244)^{1/3}\nu^{1/6}v^{-1/2}D^{1/3}x^{1/2}$
with $v$ the unperturbed liquid velocity and $x$ the distance from the edge of the solid
\citep{Jousse}.
We have identified $x$ with the longest size of the dissolving crystals and
we have taken $v=\alpha v_{\mathrm{stirrer}}$ where $\alpha$, strongly dependant
on the geometry, was chosen as 0.3.
For shaken tubes experiments, the same expression was used with $v=v_{\mathrm{tube}}$.
For flume experiments, this expression was also used with the downstream velocity
of water for $v$.

At this stage we have plotted in Fig. \ref{litedelta} the inverse of the slope
of the experimental $R(c)$ curves in Fig. \ref{liteRc} for each
experiment against the DBL thickness $\delta$.
Despite the uncertainty in the determination of $\delta$ for each experimental device,
the resulting curve is linear, as predicted by Eq. \ref{R_c}.
So each experiment, according to its geometry and flowing configuration,
probes a particular value of the boundary layer thickness \footnote{We may just notice
that the gap between \cite{Raines} measurements and the fitted law seems to
corroborate the doubt shed by \cite{Dreybrodt} on this result in their
Comment.}.
Thus the discrepancies between the dissolution rate constants in the literature stem
mainly from the hydrodynamic differences between the experimental setups.

\section{DISCUSSION}
\label{discussion}

\begin{figure}
\begin{center}
\includegraphics[width=0.8\linewidth]{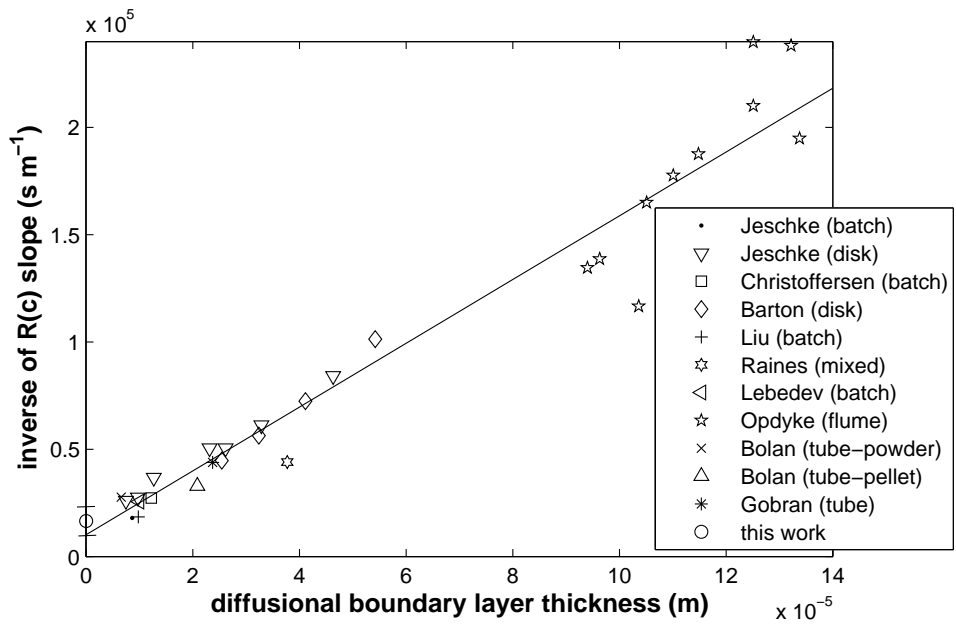}
\caption{Inverse of the $R(c)$ slopes of Fig. \ref{liteRc} as a function
of the diffusional boundary layer thickness $\delta$ for each setup.}
\label{litedelta}
\end{center}
\end{figure}

Now that a consistency has been recovered between the literature results, is it possible
to get from these results a reliable pure surface raction rate constant, comparable to our
holographic value?
The extrapolation in Fig. \ref{litedelta} of a linear curve fitting
all the literature data to $\delta=0$
gives $sc_{\mathrm{sat}}/(s_{\mathrm{r}}k_{\mathrm{s}})$.
To obtain $k_{\mathrm{s}}$ from this value, the roughness factor $\xi=s_{\mathrm{r}}/s$
is required, so we have to evaluate the mean reactive surface area $s_{\mathrm{r}}$
of the crushed samples used in the literature experiments.

Two topological classes of objects are known to have a strong influence on
$s_{\mathrm{r}}$: etch pits and steps/kinks.
For the first ones,
we benefit from a recent thorough investigation by Atomic Force Spectroscopy
of the cleavage surface of gypsum during dissolution by \cite{Fan}.
The main inference drawn from these experiments is the minor role of etch pits,
remaining always shallow, in the process.
The latter is governed by the fast dissolution of very unstable steps,
whatever the solution undersaturation.
This feature tends to enhance the similarity of gypsum with other minerals,
where etch pits seem to contribute to dissolution not via the increase of reactive
surface area but as source of monolayer steps inducing overall dissolution \citep{Luttge,Gautier}.
For the second ones, we have too few informations at the moment
about the behaviour of stepped and kinked faces of gypsum during dissolution
to estimate their influence.
So taking into account the AFM study we consider that the whole surface ---for example
measured by a BET adsorption experiment--- is reactive, i.e., $s_{\mathrm{r}}=s_{\mathrm{BET}}$.
Only one paper gives both its BET and geometric surface area values \citep{Jeschke}.
As the roughness factor $\xi=s_{\mathrm{r}}/s$ should be quite similar for all crushed gypsum
samples, we have used the value of $\xi=1100/60$ deduced from this work
for all literature experiments.
We have thereby computed $k_{\mathrm{s}}=(\partial R/\partial c)_{\delta=0} c_{\text{sat}}/\xi$
and found
$k_{\mathrm{s}}=7\times 10^{-5}$ mol m$^{-2}$ s$^{-1}$, which is the pure dissolution rate constant.
This value agrees fairly well with our $k_{\mathrm{s}}=(5\pm 2)\times 10^{-5}$ mol m$^{-2}$ s$^{-1}$
holographic value.
We can say that this coefficient had never been measured but can be deduced
from the literature measurements a posteriori, and that it
is much lower than all the rate constants proposed by the authors from their
$R(c)$ curves.

In a more graphical way, the resulting $c_{\mathrm{sat}}/(\xi k_{\mathrm{s}})$
of our holographic value has been added in Fig. \ref{litedelta} at $\delta=0$.
Remembering the experimental variability of $k_{\mathrm{s}}$, the dispersion
of the literature values and the imprecision of the roughness factor $\xi$,
the agreement between the value of $k_{\mathrm{s}}$ extrapolated from the literature results
and the value deduced from the holographic interferometry measurements is striking
and validates both our analysis of the solution chemistry experiments and
the accuracy of $k_{\mathrm{s}}$.

\section{CONCLUSION}

We have shown that despite apparent discrepancies, the correct consideration
of the hydrodynamic conditions of the global dissolution experiments in literature brings
a consistent view of the dissolution kinetics of gypsum in water and
a pure surface reaction rate constant can be computed from all these measurements,
much lower from what was expected.
We have parallely measured by holographic interferometry this pure surface
reaction coefficient in water at rest for various gypsum origins and
surface morphologies.
These two values agree well, thus validating a method
able to provide benchmark values of dissolution rate constants of minerals.

This work has two methodological consequences.
From a theoretical point of view, the holointerferometric tool
may contribute to the evaluation of models.
For instance, it permits to access directly to the prefactor $K$
in the kinetic reaction law $R=K\exp{[-\Delta G/(RT)]}$ of the transition state theory or the
dissolution plateau $A$ of the more elaborated model of \cite{Lasaga01}
of the dissolution of minerals,
these both terms being merely equal to $k_{\mathrm{s}}$.
From an experimental point of view,
the underestimation of the transport contribution to the dissolution rate can help
to explain the discrepancy between dissolution rates measured by local and global setups,
for instance the very low value of the dissolution rate of calcite and dolomite in water measured
by normal retreat of the surface \textcolor{green}{via VSI}, compared to values measured
in bulk solution chemistry experiments \citep{Arvidson,Luttge,Luttge06}.

\vspace{3ex}

%\newpage

\small
\noindent
\textit{Acknowledgements---}
We thank Elisabeth Charlaix and Jacques Bert for fruitful discussions,
Nicolas Sanchez for experimental help, and Lucienne Jardin, Sylvain Meille
and Pierre Monchoux for gypsum samples.
This work was supported by Lafarge Centre de Recherche, R\'egion Rh\^one-Alpes and CNES
(french spatial agency).

%\bibliography{bib_gypse}

\begin{thebibliography}{}

\bibitem[Anderson and Archer(2002)Anderson and Archer]{Anderson}
Anderson, D. and Archer, D. (2002).
\newblock Glacial-–interglacial stability of ocean p{H} inferred from
  foraminifer dissolution rates.
\newblock {\em Nature\/}, {\bf 416}, 70.

\bibitem[Arvidson {\em et~al.}(2003)Arvidson, Ertan, Amonette, and
  Luttge]{Arvidson}
Arvidson, R., Ertan, I., Amonette, J., and Luttge, A. (2003).
\newblock Variation in calcite dissolution rates: A fundamental problem?
\newblock {\em Geochim. Cosmochim. Acta\/}, {\bf 67}, 1623.

\bibitem[Barton and Wilde(1971)Barton and Wilde]{Barton}
Barton, A. and Wilde, N. (1971).
\newblock Dissolution rates of polycristalline samples of gypsum and
  orthorhombic forms of calcium sulphate by a rotating disc method.
\newblock {\em Trans. Faraday Soc.}, {\bf 67}, 3590.

\bibitem[Bolan {\em et~al.}(1991)Bolan, Syers, and Sumner]{Bolan}
Bolan, N., Syers, J., and Sumner, M. (1991).
\newblock Dissolution of various sources of gypsum in aqueous solutions in
  soil.
\newblock {\em J. Sci. Food Agric.}, {\bf 57}, 527.

\bibitem[Christoffersen and Christoffersen(1976)Christoffersen and
  Christoffersen]{Christoffersen}
Christoffersen, J. and Christoffersen, M. (1976).
\newblock The kinetic of dissolution of calcium sulfate dihydrate in water.
\newblock {\em J. Crystal Growth\/}, {\bf 35}, 79.

\bibitem[Colombani and Bert(2007)Colombani and Bert]{Colombani07}
Colombani, J. and Bert, J. (2007).
\newblock Holographic interferometry study of the dissolution and diffusion of
  gypsum in water.
\newblock {\em Geochim. Cosmochim. Acta\/}, {\bf 71}, 1913.

\bibitem[Dreybrodt and Gabrovsek(2000)Dreybrodt and Gabrovsek]{Dreybrodt}
Dreybrodt, W. and Gabrovsek, F. (2000).
\newblock Comments on: {M}ixed transport / reaction control of gypsum
  dissolution kinetics in aqueous solutions and initiation of gypsum karst by
  {M}ichael {A}. {R}aines and {T}homas {A}. {D}ewers in {C}hemical {G}eology
  \textbf{140}, 29-48, 1997.
\newblock {\em Chem. Geol.}, {\bf 168}, 169.

\bibitem[Fan and Teng(2007)Fan and Teng]{Fan}
Fan, C. and Teng, H. (2007).
\newblock Surface behavior of gypsum during dissolution.
\newblock {\em Chem. Geol.}, {\bf 245}, 242.

\bibitem[Gautier {\em et~al.}(2001)Gautier, Oelkers, and Schott]{Gautier}
Gautier, J., Oelkers, E., and Schott, J. (2001).
\newblock Are quartz dissolution rates proportional to {B}.{E}.{T}. surface
  areas?
\newblock {\em Geochim. Cosmochim. Acta\/}, {\bf 65}, 1059.

\bibitem[Gobran and Miyamoto(1985)Gobran and Miyamoto]{Gobran}
Gobran, G. and Miyamoto, S. (1985).
\newblock Dissolution rate of gypsum in aqueous salt solutions.
\newblock {\em Soil Sci.}, {\bf 140}, 89.

\bibitem[Hales and Emerson(1997)Hales and Emerson]{Hales}
Hales, B. and Emerson, S. (1997).
\newblock Evidence in support of first-order dissolution kinetics of calcite in
  seawater.
\newblock {\em Earth Planet. Sci. Lett.}, {\bf 148}, 317.

\bibitem[Jeschke {\em et~al.}(2001)Jeschke, Vosbeck, and Dreybrodt]{Jeschke}
Jeschke, A., Vosbeck, K., and Dreybrodt, W. (2001).
\newblock Surface controlled dissolution rates of gypsum in aqueous solutions
  exhibit nonlinear dissolution kinetics.
\newblock {\em Geochim. Cosmochim. Acta\/}, {\bf 65}, 27.

\bibitem[Jordan {\em et~al.}(2007)Jordan, Pokrovsky, Guichet, and
  Schmahl]{Jordan}
Jordan, G., Pokrovsky, O., Guichet, X., and Schmahl, W. (2007).
\newblock Organic and inorganic ligand effects on magnesite dissolution at 100
  $^{\circ}$c and ph=5 to 10.
\newblock {\em Chem. Geol.}, {\bf 242}, 484.

\bibitem[Jousse {\em et~al.}(2005)Jousse, Jongen, and Agterof]{Jousse}
Jousse, F., Jongen, T., and Agterof, W. (2005).
\newblock A method to dynamically estimate the diffusion boundary layer from
  local velocity conditions in laminar flows.
\newblock {\em Int. J. Heat Mass Transfer\/}, {\bf 48}, 1563.

\bibitem[Kemper {\em et~al.}(1975)Kemper, Olsen, and DeMooy]{Kemper}
Kemper, W., Olsen, J., and DeMooy, C. (1975).
\newblock Dissolution rate of gypsum in flowing water.
\newblock {\em Soil Sci. Soc. Am. Proc.}, {\bf 39}, 458.

\bibitem[Keren and O'Connor(1982)Keren and O'Connor]{Keren}
Keren, R. and O'Connor, G. (1982).
\newblock Gypsum dissolution and sodic soil reclamation as affected by water
  flow velocity.
\newblock {\em Soil Sci. Soc. Am. J.}, {\bf 46}, 26.

\bibitem[Lasaga(1998)Lasaga]{Lasaga98}
Lasaga, A. (1998).
\newblock {\em Kinetic theory in the earth sciences\/}.
\newblock Princeton University Press, Princeton.

\bibitem[Lasaga and Luttge(2001)Lasaga and Luttge]{Lasaga01}
Lasaga, A. and Luttge, A. (2001).
\newblock Variation in crystal dissolution rate based on a dissolution stepwave
  model.
\newblock {\em Science\/}, {\bf 291}, 2400.

\bibitem[Lebedev and Lekov(1990)Lebedev and Lekov]{Lebedev}
Lebedev, A. and Lekov, A. (1990).
\newblock Dissolution kinetics of natural-gypsum in water at 5-25$^{\circ}${C}.
\newblock {\em Geochem. Int.}, {\bf 27}, 85.

\bibitem[Liu and Nancollas(1971)Liu and Nancollas]{Liu}
Liu, S. and Nancollas, G. (1971).
\newblock The kinetic of dissolution of calcium sulfate dihydrate.
\newblock {\em J. Inorg. Nucl. Chem.}, {\bf 33}, 2311.

\bibitem[Luttge(2006)Luttge]{Luttge06}
Luttge, A. (2006).
\newblock Crystal dissolution kinetics and gibbs free energy.
\newblock {\em J. Electron Spectrosc. Rel. Phenom.}, {\bf 150}, 248.

\bibitem[Luttge {\em et~al.}(2003)Luttge, Winkler, and Lasaga]{Luttge}
Luttge, A., Winkler, U., and Lasaga, A. (2003).
\newblock Interferometric study of the dolomite dissolution: A new conceptual
  model for mineral dissolution.
\newblock {\em Geochim. Cosmochim. Acta\/}, {\bf 67}, 1099.

\bibitem[Morse {\em et~al.}(2007)Morse, Arvidson, and Luttge]{Morse}
Morse, J., Arvidson, R., and Luttge, A. (2007).
\newblock Calcium carbonate formation and dissolution.
\newblock {\em Chem. Rev.}, {\bf 107}, 342.

\bibitem[Opdyke {\em et~al.}(1987)Opdyke, Gust, and Ledwell]{Opdyke}
Opdyke, B., Gust, G., and Ledwell, J. (1987).
\newblock Mass transfer from smooth alabaster surfaces in turbulent flows.
\newblock {\em Geophys. Res. Lett.}, {\bf 14}, 1131.

\bibitem[Porter {\em et~al.}(2000)Porter, Sanford, and Suttles]{Porter}
Porter, E., Sanford, L., and Suttles, S. (2000).
\newblock Gypsum dissolution is not a universal integrator of 'water motion'.
\newblock {\em Limnol. Oceanogr.}, {\bf 45}, 145.

\bibitem[Raines and Dewers(1997)Raines and Dewers]{Raines}
Raines, M. and Dewers, T. (1997).
\newblock Mixed transport / reaction control of gypsum dissolution kinetics in
  aqueous solutions and initiation of gypsum karst.
\newblock {\em Chem. Geol.}, {\bf 140}, 29.

\bibitem[Raju and Atkinson(1990)Raju and Atkinson]{Raju}
Raju, K. and Atkinson, G. (1990).
\newblock The thermodynamics of "scale" mineral solubilities. 3. {C}alcium
  sulfate in aqueous {N}a{C}l.
\newblock {\em J. Chem. Eng. Data\/}, {\bf 35}, 361.

\bibitem[Rickard and Sj\"oberg(1983)Rickard and Sj\"oberg]{Rickard}
Rickard, D. and Sj\"oberg, E. (1983).
\newblock Mixed kinetic control of calcite dissolution rates.
\newblock {\em Am. J. Sci.}, {\bf 283}, 815.

\bibitem[Vinson and Luttge(2005)Vinson and Luttge]{Vinson}
Vinson, M. and Luttge, A. (2005).
\newblock Multiple length-scale kinetics: an integrated study of calcite
  dissolution rates and strontium inhibition.
\newblock {\em Am. J. Sci.}, {\bf 305}, 119.

\end{thebibliography}
%\bibliographystyle{natbib}

%\section*{List of figures}

%\clearpage

\end{document}